\documentclass[aps,prb,twocolumn,10pt, showkeys,showpacs,amsmath,amsfonts,floatfix,superscriptaddress,nofootinbib,longbibliography]{revtex4-2}

\usepackage{chemformula}

\usepackage{amssymb,amsmath,amstext}                
\usepackage{graphicx}                                               
\usepackage{epstopdf}                                               
\usepackage{color}       
\usepackage{bm}
\usepackage{blindtext}
\usepackage{appendix}                                              
\usepackage[utf8]{inputenc}
\usepackage{ulem}
\usepackage{latexsym}

\usepackage[colorlinks=true,citecolor=blue,linkcolor=magenta]{hyperref}
\def\be{\begin{equation}}
\def\ee{\end{equation}}
\def\bea{\begin{eqnarray}}
\def\eea{\end{eqnarray}}
\def\bi{\begin{itemize}}
\def\ei{\end{itemize}}

\newcommand{\ket}[1]{\mbox{$| #1 \rangle$}}

\makeatletter
\def\paragraph{%
  \@startsection
    {paragraph}%
    {4}%
    {\parindent}%
    {\z@}%
    {-0em}%
    {\normalfont\normalsize\itshape}%
}%
\makeatother

\begin{document}

\title{Variational boundary based tensor network renormalization group}

\author{Feng-Feng Song}
\email{song@issp.u-tokyo.ac.jp}
\affiliation{Institute for Solid State Physics, The University of Tokyo, Kashiwa, Chiba 277-8581, Japan}

\author{Naoki Kawashima}
\email{kawashima@issp.u-tokyo.ac.jp}
\affiliation{Institute for Solid State Physics, The University of Tokyo, Kashiwa, Chiba 277-8581, Japan}
\affiliation{Trans-scale Quantum Science Institute, The University of Tokyo, Bunkyo, Tokyo 113-0033, Japan}

\date{\today}

\begin{abstract}
We propose a real-space renormalization group algorithm for accurately coarse-graining two-dimensional tensor networks. The central innovation of our method lies in utilizing variational boundary tensors as a globally optimized environment for the entire system. Based on this optimized environment, we construct renormalization projectors that significantly enhance accuracy. By leveraging the canonical form of tensors, our algorithm maintains the same computational complexity as the original tensor renormalization group (TRG) method, yet achieves higher accuracy than existing approaches that do not incorporate entanglement filtering. Our work offers a practical pathway for extending TRG methods to higher dimensions while keeping computational costs manageable.
\end{abstract}

\maketitle

\section{Introduction}
Tensor network methods have emerged as powerful tools for the theoretical and numerical study of strongly correlated many-body systems in both classical and quantum systems~\cite{Orus2014,Haegeman2017,Orus2019}. By representing partition functions or quantum wavefunctions as networks of locally connected tensors, these methods enable efficient encoding of exponentially large configuration spaces. A central computational challenge in this framework is the contraction of large-scale tensor networks, which becomes increasingly demanding with system size. To address this, two major algorithmic strategies have been developed: (i) boundary methods, such as the density matrix renormalization group (DMRG)~\cite{White1992, White1993, Nishino1995}, the corner transfer matrix renormalization group (CTMRG)~\cite{Baxter1978,Nishino1996,Nishino1997} and the time evolving block decimation (TEBD) algorithm~\cite{Vidal2003,Vidal2007classical}, which approximate the dominant eigenvectors of the transfer matrix; and (ii) real-space renormalization group (RG) methods~\cite{Xiang2023}, which employ coarse-graining transformations to systematically reduce the degrees of freedom while preserving long-wavelength physics. The latter approach gives rise to a renormalization flow of local tensors and has proven particularly useful for studying universal properties and critical phenomena~\cite{Gu2009,Lyu2021}.

Among real-space RG techniques, the tensor renormalization group (TRG) method, initially introduced by Levin and Nave~\cite{Levin2007}, provides a systematic scheme for coarse-graining two-dimensional (2D) tensor networks using singular value decomposition (SVD). At each renormalization step, TRG merges neighboring tensors and performs SVD to truncate smaller singular values, thereby reducing bond dimensions and controlling the computational cost. This iterative procedure enables approximate evaluation of the partition function in the thermodynamic limit. Since its introduction, TRG has served as a foundation for a broad class of tensor network coarse-graining algorithms. However, it is now well understood that TRG exhibits two key limitations, particularly in the vicinity of criticality. First, TRG does not effectively remove short-range correlations during coarse-graining, which leads to the accumulation of irrelevant local structures and a breakdown of scale invariance~\cite{Gu2009,Ferris2013}. Second, the truncation step in TRG is based solely on local information, especially the singular values of a single tensor pair, and therefore does not yield an optimal approximation from the perspective of the entire tensor network. These limitations result in increasing truncation errors under iteration and reduced accuracy near critical points. 

To address the limitations of TRG, a wide range of extensions have been proposed, which can be broadly categorized into two approaches: those that improve the truncation scheme by incorporating environmental effects, and those that aim to remove local short-range correlations through entanglement filtering. The first category focuses on optimizing the truncation process by including information from the surrounding tensor environment, thereby yielding more globally informed approximations. Representative methods include the second renormalization group (SRG)~\cite{Xie2009,Zhao2010}, higher-order TRG (HOTRG)~\cite{Xie2012,Garcia-Saez2013}, higher-order SRG (HOSRG)~\cite{Zhao2016,Chen2020}, CTMRG-based boundary TRG~\cite{Iino2019}, CTM-TRG~\cite{Morita2021}, and the bond-weighted TRG (BWTRG)~\cite{Adachi2022}. The second category targets the explicit removal of redundant local structures that obscure long-range entanglement, such as short-range loops. This is achieved by introducing disentangling transformations or filtering procedures~\cite{Vidal2007,Evenbly2009,Gu2009,Bal2017,Evenbly2017implicitly,Ying2017,Harada2018,Evenbly2018,Lee2020}. Notable examples include tensor network renormalization (TNR)~\cite{Evenbly2015,Evenbly2017}, loop-TRG~\cite{Yang2017} and its variant, nuclear norm regularization TRG (NNR-TRG)~\cite{Homma2024}, and graph-independent local truncation (GILT)~\cite{Hauru2018}. Although these approaches significantly improve the accuracy, they often come with increased computational complexity due to the use of non-local updates and more sophisticated tensor manipulations.

In this work, we introduce a variational boundary-based tensor network renormalization group (VBTRG) method that significantly improves the accuracy of the TRG for 2D tensor networks. The overall structure of VBTRG closely follows that of the HOTRG, where local tensors are successively coarse-grained through a sequence of bond-merging projections. However, the key innovation lies in how these projection operators are determined. Unlike conventional HOTRG, which constructs local projectors based on truncated SVD, VBTRG employs variational boundary matrix product states (MPS) to approximate the global environment of the infinite system. Using this global information, VBTRG optimizes the bond-merging projectors with high precision, leading to significantly improved accuracy. 
This global optimization strategy preserves the computational complexity of the original TRG method while outperforming existing environment-optimized methods in accuracy. Our method, in contrast to most other high performance methods mentioned above, does not remove the redundant loops, while its performance is better than or equally good as them. Moreover, the scalability of VBTRG offers a promising foundation for extending tensor network renormalization to higher-dimensional systems.

\section{Methods}

\subsection{Variational boundary tensors}
\begin{figure}[tbp]
    \centering
    \includegraphics[width=0.99\linewidth]{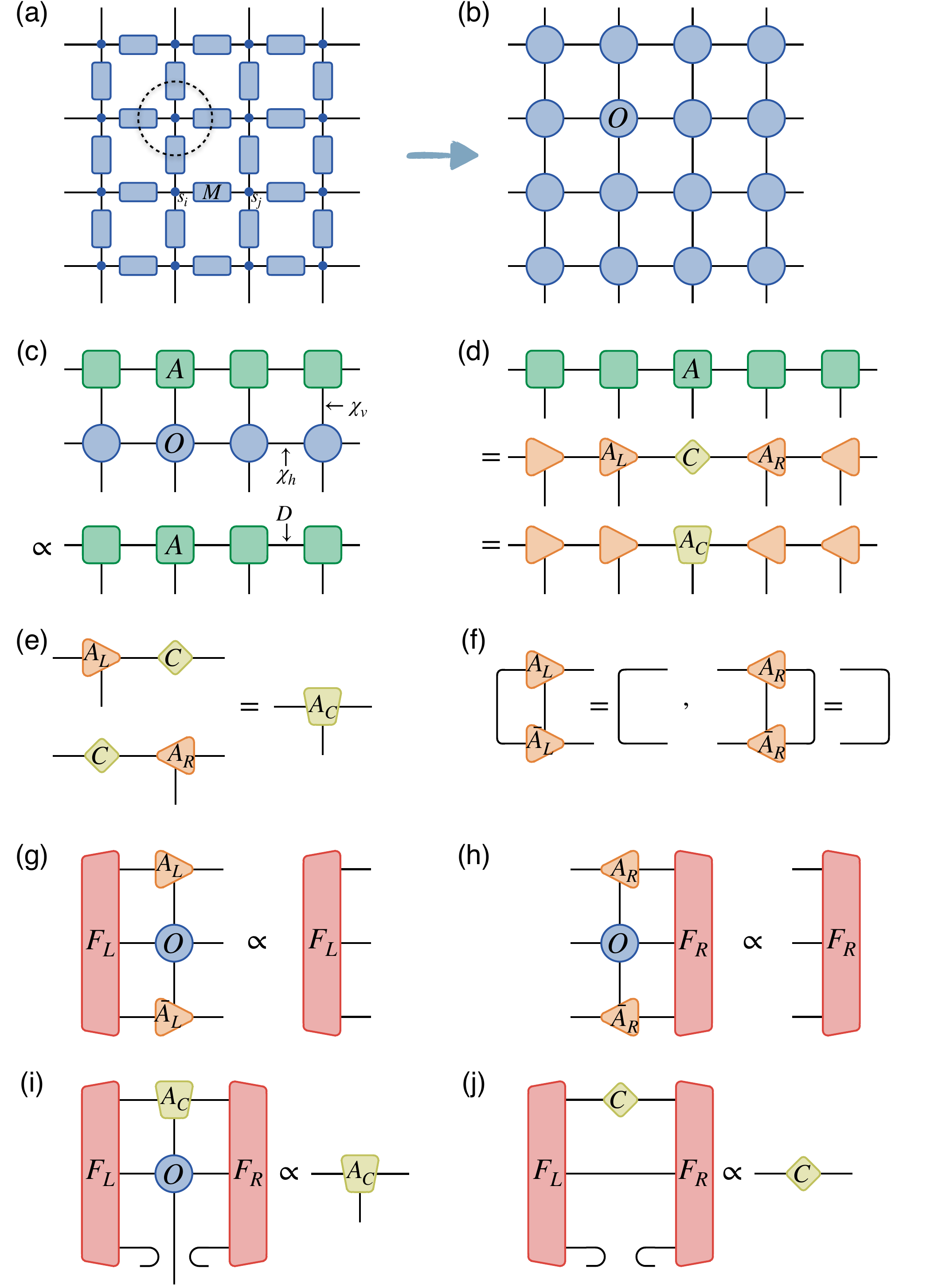}
    \caption{ 
    (a) The tensor network representation of the partition function with interaction matrices on the links accounting for the Boltzmann weight. 
    (b) The infinite tensor network representation of the partition function composed of uniform local $O$ tensors.
    (c) Eigenequation for the fixed-point MPS $\ket{\Psi(A)}$ of the row-to-row transfer operator $\mathcal{T}$. 
    (d) The uniform representation and two equivalent mixed canonical forms of the fixed-point MPS. 
    (e) The isometric gauge transformation between the left canonical tensor $A_L$ and the right canonical $A_R$. 
    (f) The canonical conditions of the fixed-point local tensors.
    (g) and (h) Eigen-equations to update the left and right environmental eigenvectors of the channel operators.
    (i) and (j) Eigen-equations to obtain the central tensors based on the new channel environment.}
    \label{fig:vumps}
\end{figure}

Within the framework of tensor network methods, the partition function of a statistical model with local interactions can be expressed as the contraction of an infinite tensor network~\cite{Zhao2010,Vanhecke2021,Song2023}. For instance, the Hamiltonian of the 2D Ising model on a square lattice is given by
\be
H = -J \sum_{\langle i,j\rangle} s_i s_j,
\ee
where each Ising spin takes values $s_i = \pm 1$, and the summation runs over all nearest-neighbor (NN) pairs $\langle i,j \rangle$. The corresponding partition function is
\be
Z = \sum_{\{s_i\}} e^{-\beta H(\{s_i\})} = \sum_{\{s_i\}} \prod_{\langle i,j \rangle} M(s_i, s_j),
\ee
where the interaction matrix $M(s_i, s_j) = e^{\beta J s_i s_j}$ is defined on each NN bond, as illustrated in Fig.~\ref{fig:vumps}(a). By factorizing the interaction matrix as $M = WW^\dagger$, with
\be
W = \begin{pmatrix}
\sqrt{\cosh(\beta J)} & \sqrt{\sinh(\beta J)} \\
\sqrt{\cosh(\beta J)} & -\sqrt{\sinh(\beta J)}
\end{pmatrix},
\ee
the partition function can be reformulated as a uniform tensor network, shown in Fig.~\ref{fig:vumps}(b), in the form
\be
Z = \mathrm{tTr}\prod_i O_{m,n,k,l}(i),
\ee
where ``tTr" denotes the tensor contraction, and each four-leg tensor $O$ of bond dimension $\chi=2$ is constructed by contracting the $W$ matrices connected to the same site
\be
O_{m,n,k,l} = \sum_{s_i = \pm 1} W(s_i, s_m) W(s_i, s_n) W(s_i, s_k) W(s_i, s_l).
\ee

Unlike TRG and HOTRG, where the bond-merging projectors are determined through optimization over local clusters, our approach seeks to obtain these projectors via global optimization. Specifically, this requires contracting the entire tensor network outside the local region where the projectors are applied. However, performing an exact contraction over such a large environment is computationally intractable, as the cost increases rapidly with system size. To address this, suitable approximations to the global environment must be employed. The HOSRG algorithm achieves this by performing forward and backward iterations to approximate the full contraction, resulting in a computational cost of $\mathcal{O}(\chi^7)$~\cite{Xie2012,Zhao2016}. Later, the CTM-TRG method improves upon this by using a corner transfer matrix (CTM) environment, composed of corner and edge tensors, reducing the cost of computing the bond-merging operators to $\mathcal{O}(\chi^6)$~\cite{Morita2021}.

Here, we adopt an alternative environment approximation based on variational boundary matrix product states (MPS), known as the variational uniform matrix product state (VUMPS) algorithm~\cite{Haegeman2017,Zauner-Stauber2018,Vanderstraeten2019tan}. VUMPS is among the most efficient methods for computing the dominant eigenvector of the row-to-row transfer matrix
\be
\mathcal{T} = \mathrm{tTr}(\cdots O\, O\, O\, O \cdots),
\ee
as depicted in Fig.~\ref{fig:vumps}(c). The corresponding fixed-point equation is
\be
\mathcal{T} \ket{\Psi(A)} = \Lambda_{\max} \ket{\Psi(A)},
\label{eq:TM}
\ee
where the leading eigenvector is represented by an infinite MPS composed of periodic three-leg tensors $A$, with auxiliary bond dimension $D$ and physical bond dimension $\chi_v$.
To facilitate efficient variational optimization, the uniform MPS is brought into its mixed canonical form, as illustrated in Fig.~\ref{fig:vumps}(d),
\begin{align}
\ket{\Psi(A)} &= \mathrm{tTr}(\cdots A_L A_L\, C\, A_R A_R \cdots) \\
              &= \mathrm{tTr}(\cdots A_L A_L\, A_C\, A_R A_R \cdots),
\end{align}
where $A_L$ and $A_R$ are the left- and right-canonical tensors, respectively, and $C$ is a central bond matrix. These tensors are related by the gauge transformation shown in Fig.~\ref{fig:vumps}(e),
\be
A_L\, C = A_C = C\, A_R,
\ee
and satisfy the isometric constraints
\be
A_L A_L^\dagger = A_R A_R^\dagger = I,
\label{eq:canonical}
\ee
as illustrated in Fig.~\ref{fig:vumps}(f).

The VUMPS algorithm proceeds by iteratively solving two sets of local eigenvalue equations. 
(i) For the left and right channel environments, we solve
\be
\mathbb{T}_L F_L \propto F_L, \quad \mathbb{T}_R F_R \propto F_R,
\label{eq:FLR}
\ee
as illustrated in Fig.~\ref{fig:vumps}(g) and (h), where $\mathbb{T}_L$ and $\mathbb{T}_R$ are the channel transfer operators constructed from the tensors $A_L$ and $A_R$, respectively.
(ii) For the central tensors $A_C$ and $C$, we solve
\be
H_{A_C} A_C \propto A_C, \quad H_C C \propto C,
\label{eq:AC}
\ee
as shown in Fig.~\ref{fig:vumps}(i) and (j), where $H_{A_C}$ and $H_C$ are the effective Hamiltonians formed from the left and right fixed-point tensors $F_L$ and $F_R$.

The VUMPS algorithm exhibits a computational complexity of $\mathcal{O}(D^3)$ and demonstrates fast convergence. As a result, the initial boundary MPS for the unrenormalized original $O$ tensors can be computed very efficiently due to the small physical bond dimension.

\subsection{Construction of the projectors}
\begin{figure}[tbp]
    \centering
    \includegraphics[width=0.99\linewidth]{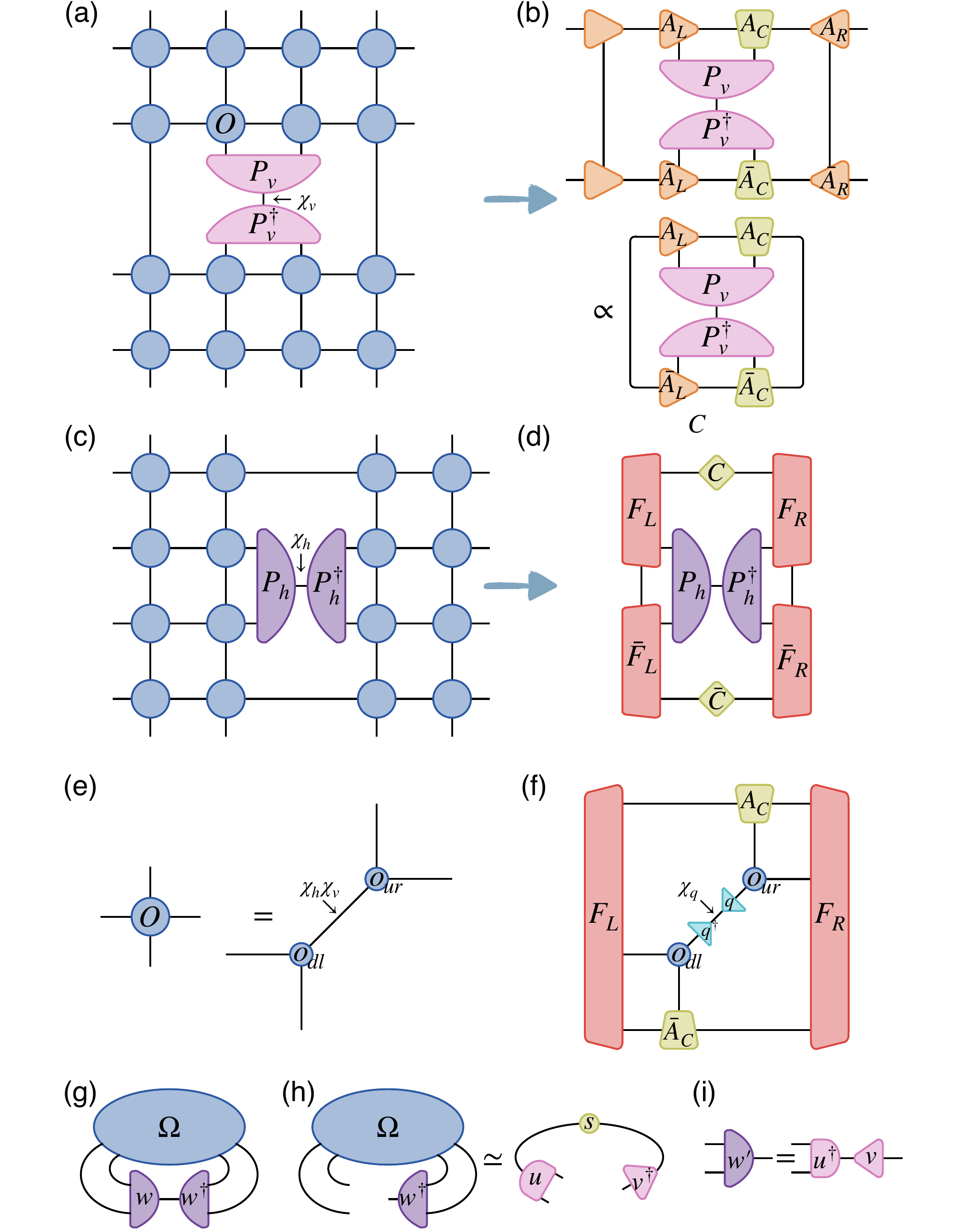}
    \caption{
    (a) The global tensor network with a pair of vertical projectors $P_v$ and $P_v^\dagger$ of truncated bond dimension $\chi_v$ inserted in. 
    (b) The optimization problem for the projectors is simplified into the minimization of the error in the contraction of a small tensor graph using the boundary MPS. 
    (c) The global tensor network with a pair of horizontal projectors $P_h$ and $P_h^\dagger$ of truncated bond dimension $\chi_h$ inserted in. 
    (d) The projectors are obtained by minimization of the error in the contraction of a small tensor graph using the fixed-point conditions. 
    (e) The local tensor $O$ is decomposed into two three-index up-right tensor $o_{ur}$ and down-left tensor $o_{dl}$ without truncation. 
    (f) An additional approximation using projectors $q$ and $q^\dagger$ to reduce the bond dimension of $o$ tensors from $\chi_h\chi_v$ to $\chi_q$, based on the global environment. 
    (g) The optimization problem for the projectors is represented as finding the isometries $w$ insert into the global environment $\Omega_w$. 
    (h) Environment $\Gamma_w=\Omega_w w^\dagger$ is decomposed via SVD, into a product of isometric tensors $u$, $v$, and diagonal matrix $s$.
    (i) The isometry $w$ is updated as $w'=u^\dagger v$ to minimize the loss.}
    \label{fig:projectors}
\end{figure}

Based on the variational boundary tensors, the computational cost of evaluating the bond-merging operator is reduced to $\mathcal{O}(\chi^5)$. The bond-merging projectors are inserted into local clusters to reduce the bond dimension of grouped tensor pairs, while minimizing the discrepancy between the full tensor network representations of the partition function with and without the inserted projectors.

The vertical bond-merging projector pair, $P_v$ and $P_v^\dagger$, is illustrated in Fig.~\ref{fig:projectors}(a), where the two vertical bonds of the $O$ tensors are truncated to a reduced bond dimension $\chi_v$. As shown in Fig.~\ref{fig:projectors}(b), using the top and bottom MPS fixed points $\ket{\Psi(A)}$ in Eq.~\eqref{eq:TM}, the full network is first compressed into an infinite one-dimensional tensor train and then further reduced to a smaller network involving only the tensors $A_L$, $A_C$, and $P_v$, by applying the isometric conditions in Eq.~\eqref{eq:canonical}.

Similarly, the infinite network containing horizontal bond-merging projectors $P_h$ and $P_h^\dagger$, shown in Fig.~\ref{fig:projectors}(c), can be reduced to a small tensor graph as depicted in Fig.~\ref{fig:projectors}(d) using the relations in Eq.~\eqref{eq:FLR} and \eqref{eq:AC}. Here the horizontal bond dimensions are merged and truncated to $\chi_h$.

Instead of directly applying the projectors $P_h$ and $P_v$ as in HOTRG, which leads to a coarse-graining cost of $\mathcal{O}(\chi^7)$, we adopt an additional approximation to reduce the bond dimension of the intermediate tensors~\cite{Morita2021}. 
As illustrated in Fig.~\ref{fig:projectors}(e), we first decompose the $O$ tensor with horizontal bond dimension $\chi_h$ and vertical bond dimension $\chi_v$ into a pair of three-leg $o$ tensors using SVD
\be
O_{mn,kl} = (U \sqrt{\Sigma})(\sqrt{\Sigma} V^\dagger) = \sum_{h=1}^{\chi_h \chi_v} o_{mn,h} \, o_{h,kl}.
\ee
Here, $U$ and $V$ are unitary matrices, and $\Sigma$ is the diagonal matrix of singular values. 
Then, a pair of truncation operators, $q$ and $q^\dagger$, with reduced bond dimension $\chi_q$, is constructed in a similar way as the bond-merging projectors, as shown in Fig.~\ref{fig:projectors}(f).

The determination of the projectors $P_h$, $P_v$, and $q$ can be unified under a common optimization framework, analogous to optimizing an isometry $w$ for its environment $\Omega_w$, by maximizing the contraction $\mathrm{tTr}(\Omega_w w w^\dagger)$, as illustrated in Fig.~\ref{fig:projectors}(g). The isometry $w$ is updated iteratively using an SVD-based approach~\cite{Evenbly2017}. 
In each iteration, we fix $w^\dagger$ and construct a temporary environment $\Gamma_w = \Omega_w w^\dagger$. As shown in Fig.~\ref{fig:projectors}(h), we then perform SVD on $\Gamma_w$, yielding $\Gamma_w = u s v^\dagger$. The isometry is subsequently updated as $w' = u^\dagger v$, as depicted in Fig.~\ref{fig:projectors}(i). Finally, the isometry is rescaled to satisfy $\mathrm{Tr}\,\Omega=\mathrm{tTr}(\Omega_w w w^\dagger)$.
It is worth noting that while directly constructing the full environment $\Omega$ for $P_h$ and $P_v$ incurs a computational cost of $\mathcal{O}(\chi^6)$, forming the intermediate environment $\Gamma_w$ requires only $\mathcal{O}(\chi^5)$. Consequently, the overall cost of obtaining the projectors scales as $\mathcal{O}(\chi^5)$.

\subsection{The renormalization process}
\begin{figure}[tbp]
    \centering
    \includegraphics[width=0.99\linewidth]{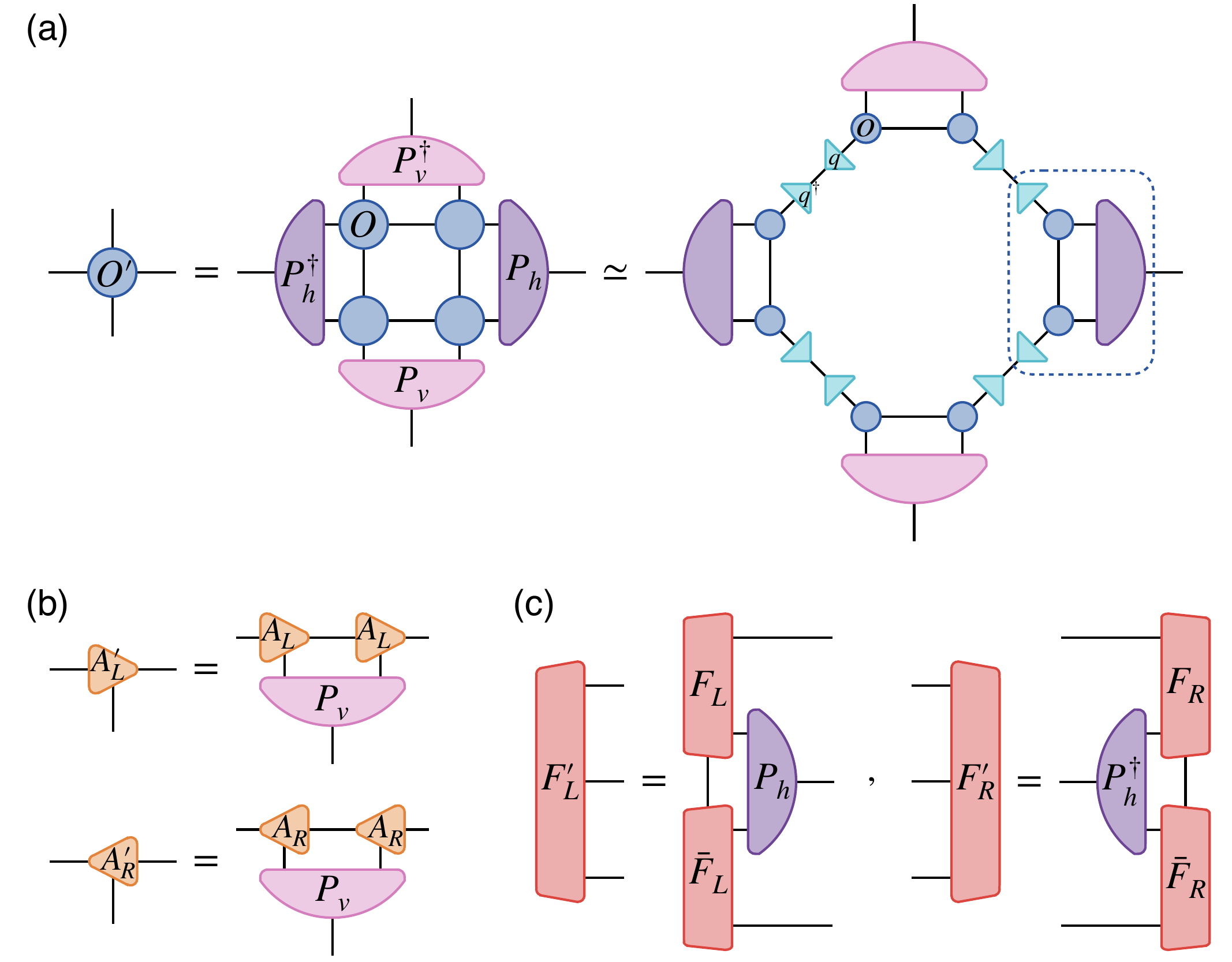}
    \caption{ 
    (a) The coarse-grained tensor $O'$ is obtained by the simultaneous projections along the horizontal and vertical directions for an $2\times2$ cluster. The computational cost of the contraction is reduced by the intermediate projectors $q$ and $q^\dagger$. After contraction of the sub-network in the dashed circle at a complexity of $\mathcal{O}(\chi^5)$, we arrive at the same structure as the original TRG.
    (b) Update of the boundary canonical tensors using the merging projectors $P_v$.
    (c) Update of the left and right channel environments using the merging projectors $P_h$.}
    \label{fig:rg}
\end{figure}

By introducing the intermediate projectors $q$, the contraction of the coarse-grained tensor from a $2 \times 2$ block can be decomposed into smaller sub-networks, as illustrated in Fig.~\ref{fig:rg}(a). The computational cost for contracting each sub-network within the dashed circles is $\mathcal{O}(\chi^5)$. After contracting the four sub-networks, the structure effectively reduces to that of the original TRG, where the updated tensor $O'$ is obtained by contracting four three-leg tensors. The overall cost of the coarse-graining step in TRG can be further reduced to $\mathcal{O}(\chi^5)$ by employing randomized SVD~\cite{Morita2018} or iterative SVD~\cite{Evenbly2017} to generate the intermediate three-leg tensors.

Meanwhile, the boundary environment can be efficiently updated using simple projection steps. As shown in Fig.~\ref{fig:rg}(b), the updated tensors $A_L'$ and $A_R'$ are obtained by merging two $A_L$ and $A_R$ tensors using the vertical projector $P_v$. Similarly, the channel environment tensors $F_L'$ and $F_R'$ are updated via the horizontal projectors $P_h$, as illustrated in Fig.~\ref{fig:rg}(c). 
While this projection-based update is computationally efficient, it is less accurate than the environment obtained from the VUMPS method applied to the coarse-grained tensor $O'$. To enhance the quality of the global environment, one can perform a single iteration of the VUMPS algorithm using the coarse-grained boundary tensors as the initial input. In practice, we find that a single VUMPS update is sufficient to produce a high-fidelity approximation of the global environment. Consequently, the cost of the environment update remains lower than that of coarse-graining $O'$.

\section{Results}
\begin{figure}[tbp]
    \centering
    \includegraphics[width=0.99\linewidth]{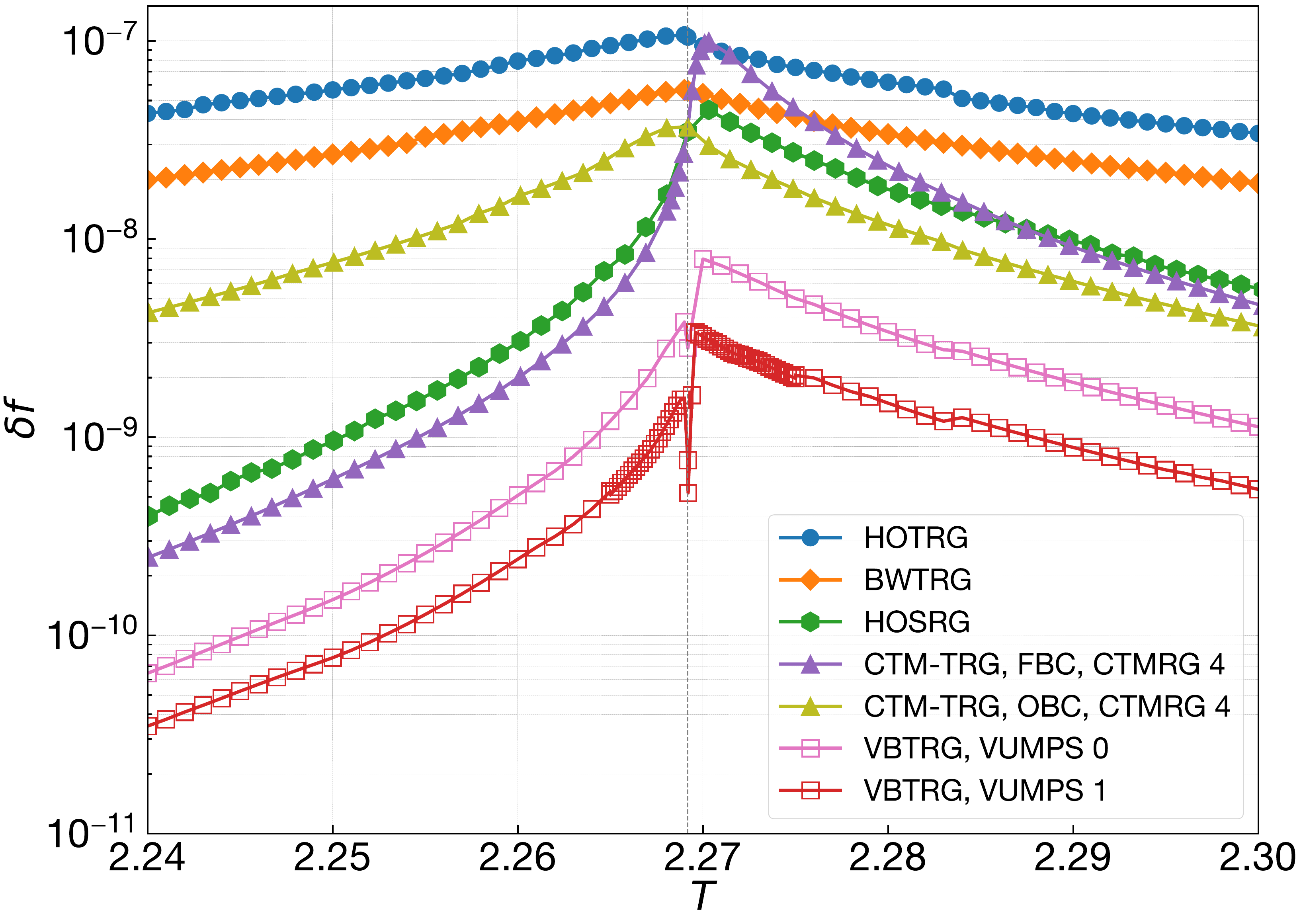}
    \caption{
    Relative errors in the free energy per site of the classical Ising model near the critical temperature, obtained using HOTRG~\cite{Xie2012}, BWTRG~\cite{Adachi2022}, HOSRG~\cite{Chen2020}, CTM-TRG~\cite{Morita2021}, and our VBTRG calculations at bond dimension $\chi = 24$. The accuracy of CTM-TRG depends on whether fixed boundary conditions (FBC) or open boundary conditions (OBC) are used for initializing the CTM environment. In each CTM-TRG update, four CTMRG steps are performed. For VBTRG, zero or one VUMPS step is applied in each update to approximate the global environment. The vertical dashed line denotes the critical temperature.}
    \label{fig:fe_T}
\end{figure}

We study the 2D Ising model on the square lattice using the VBTRG method and compare its performance with other state-of-the-art approaches that focus on truncation environment optimization, including HOTRG~\cite{Xie2012}, BWTRG~\cite{Adachi2022}, HOSRG~\cite{Chen2020}, and CTM-TRG~\cite{Morita2021}. The initial boundary tensors required before the renormalization process are efficiently obtained using VUMPS, with minimal computational overhead, as the converged boundary tensors at one temperature can be reused as initial inputs for neighboring temperatures~\cite{Zauner-Stauber2018}. Similar to CTM-TRG, the VBTRG method exhibits rapid convergence toward the thermodynamic limit within a few RG steps, due to its use of the global boundary environment~\cite{Morita2021}. In our implementation, we perform 20 renormalization steps, with the truncation bond dimensions of the boundary tensors and merging projectors set to $D=\chi_h = \chi_v = \chi$, and the bond dimension of the intermediate projector $q$ set to $\chi_q = 2\chi$.

The temperature dependence of the relative error in the free energy per site,
\begin{equation}
\delta f = \left| \frac{f - f_{\text{exact}}}{f_{\text{exact}}} \right|,
\end{equation}
is shown in Fig.~\ref{fig:fe_T}. All TRG-based methods are compared using a fixed truncation bond dimension of $\chi = 24$. It is evident that VBTRG consistently yields lower relative errors than the other methods across the entire temperature range, even without intermediate updates of the boundary tensors using VUMPS (denoted as ``VUMPS 0" in the figure). The accuracy improves further when one VUMPS update is performed at each renormalization step (labeled ``VUMPS 1"). While a single VUMPS iteration may not achieve full convergence, it is sufficient to provide a high-quality global environment. Additional VUMPS iterations beyond the first yield only marginal improvements in accuracy. A slight improvement in accuracy is observed near the critical temperature, $T_c = 2/\ln(1 + \sqrt{2})$, likely due to the higher quality of the initial boundary environment. This results from the fixed convergence criterion in the VUMPS initialization for the boundary tensors, which demands more iterations near criticality.

\begin{figure}[tbp]
    \centering
    \includegraphics[width=0.99\linewidth]{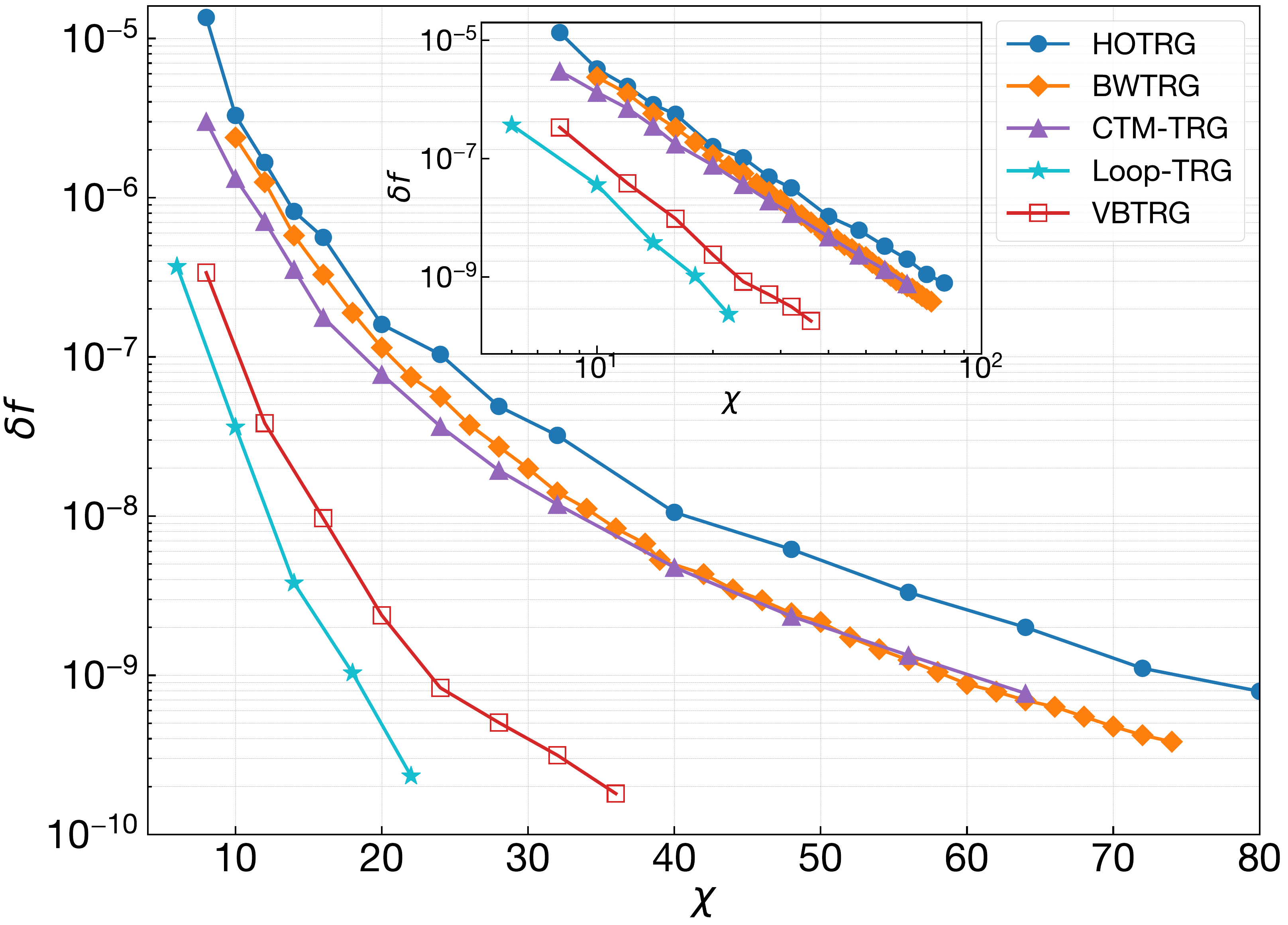}
    \caption{
    Bond dimension dependence of the relative error of the free energy per site at the critical temperature.}
    \label{fig:fe_D}
\end{figure}

Furthermore, we compare the bond-dimension dependence of the relative error in the free energy at the critical temperature. As shown in Fig.~\ref{fig:fe_D}, VBTRG achieves significantly lower relative errors across all bond dimensions than other methods that focus on optimizing the truncation environment. Remarkably, its accuracy approaches that of state-of-the-art Loop-TRG methods, which combine entanglement filtering with variational optimization~\cite{Yang2017,Homma2024}, although a small gap in precision still remains.

\section{Conclusion and outlook}
In conclusion, we introduce a variational boundary-based tensor network renormalization group (VBTRG) method that significantly improves the accuracy of coarse-graining two-dimensional (2D) tensor networks. By employing variational boundary matrix product states (MPS) to approximate the global environment, VBTRG optimizes the truncation process while maintaining a computational complexity comparable to the original TRG method, with a potential scaling of $\mathcal{O}(\chi^5)$. The use of variational boundary tensors enables a high-fidelity representation of the global environment, leading to consistently improved accuracy over existing methods, including HOTRG and HOSRG with $\mathcal{O}(\chi^7)$~\cite{Xie2012,Zhao2016,Chen2020}, CTM-TRG with $\mathcal{O}(\chi^6)$~\cite{Morita2021}, and BWTRG with $\mathcal{O}(\chi^5)$~\cite{Adachi2022}. Benchmark results for the 2D Ising model show that VBTRG yields significantly lower errors in the free energy across all temperatures and bond dimensions. Its precision even gets close to that of state-of-the-art Loop-TRG methods~\cite{Yang2017,Homma2024}, which incorporate entanglement filtering and variational optimization.

The advantage of VBTRG stems from the rapid convergence of boundary tensors in the VUMPS algorithm. In contrast to the CTM environment used in CTM-TRG, based on the standard CTMRG method~\cite{Corboz2014}, the variational boundary tensors in VUMPS adapt more flexibly to both ferromagnetic and paramagnetic phases. While VUMPS is most effective for Hermitian transfer operators, it has been readily extended to other variationally inspired methods, such as the biorthonormal transfer-matrix renormalization group (BTMRG)~\cite{Huang2011,Tang2025}, the improved fixed-point corner method (FPCM)~\cite{Fishman2018}, multi-unitcell VUMPS~\cite{Nietner2020}, and applications on general lattice geometries~\cite{Nyckees2023}. These variational boundary approaches provide a promising pathway to extend our method to more complex systems.

The VBTRG framework also offers a promising route for extending tensor network renormalization to higher-dimensional systems. Its efficient use of variational boundary tensors, which typically have lower ranks, may help reduce the computational cost associated with obtaining bond-merging operators and performing higher-order contractions. For instance, in the case of 3D tensor networks, the bond-merging projectors can be formulated using variational projected entangled-pair states (PEPS)~\cite{Verstraete2004,Nishio2004,Jordan2008}. 

Moreover, incorporating entanglement filtering techniques into VBTRG could further enhance its accuracy. Although global optimization is applied in VBTRG, short-range correlations are not fully removed. As a result, the scaling dimensions and central charges extracted from coarse-grained tensors remain unstable, similar to TRG and HOTRG. Techniques such as loop entanglement filtering~\cite{Yang2017,Homma2024}, implicit entanglement filtering~\cite{Evenbly2017implicitly}, and full environment truncation~\cite{Evenbly2018} can be readily integrated into our framework, using the global environment structures shown in Fig.~\ref{fig:projectors}(b), (d), and (f). We expect that combining a global environment with entanglement filtering will significantly enhance the accuracy and robustness of existing TRG algorithms, potentially pushing their performance to a new level.

\begin{acknowledgments}
We would like to thank Satoshi Morita for insightful discussions and for generously providing the CTM-TRG data used in this work.
This work is supported by JSPS KAKENHI (Grant No.\ 23K25789). 
FF.S. acknowledges support by  JSPS Grant-in-Aid for Early-Career Scientists (Grant No.\ 25K17311).
Part of the computation in this work has been done using the facilities of the Supercomputer Center, the Institute for Solid State Physics, the University of Tokyo.
\end{acknowledgments}

\bibliographystyle{apsrev4-2}
\bibliography{ref}
\end{document}